\documentclass[]{aastex631}

\begin{document}

\title{The Gravitational Wave AfterglowPy Analysis (GWAPA) webtool}

\correspondingauthor{R. A. J. Eyles-Ferris}
\email{raje1@leicester.ac.uk}

\author[0000-0002-8775-2365]{R. A. J. Eyles-Ferris}
\affiliation{School of Physics and Astronomy, University of Leicester \\
University Road \\
Leicester, LE1 7RH, UK}

\author{H. van Eerten}
\affiliation{Department of Physics, University of Bath \\
Claverton Down \\
Bath, BA2 7AY , UK}

\author[0000-0002-1869-7817]{E. Troja}
\affiliation{Dipartimento di Fisica, Universit\`a Degli Studi di Roma - Tor Vergata \\
via della Ricera Scientifica 1 \\
00100 Rome, IT}

\author[0000-0002-5128-1899]{P. T. O'Brien}
\affiliation{School of Physics and Astronomy, University of Leicester \\
University Road \\
Leicester, LE1 7RH, UK}



\begin{abstract}

We present the first release of the Gravitational Wave AfterglowPy Analysis (GWAPA) webtool\footnote{Available at \url{https://gwapa.web.roma2.infn.it/}}. GWAPA is designed to provide the community with an interactive tool for rapid analysis of gravitational wave afterglow counterparts and can be extended to the general case of gamma-ray burst afterglows seen at different angles. It is based on the \texttt{afterglowpy} package and allows users to upload observational data and vary afterglow parameters to infer the properties of the explosion. Multiple jet structures, including top hat, Gaussian and power laws, in addition to a spherical outflow model are implemented. A \texttt{Python} script for MCMC fitting is also available to download, with initial guesses taken from GWAPA.

\end{abstract}



\section{Introduction} \label{sec:intro}

The Laser Interferometer Gravitational-Wave Observatory (LIGO), Virgo interferometer and Kamioka Gravitational Wave Detector (KAGRA) have begun their fourth observing run (O4), enabling us to search for electromagnetic (EM) counterparts of gravitational wave sources. During O4, it is crucial for the community to have tools to rapidly respond to and analyse possible EM counterparts. To that end, we introduce the first release of the Gravitational Wave AfterglowPy Analysis (GWAPA \texttt{v0.9.17b}) webtool.

This first release of GWAPA is designed to allow users to easily analyse the afterglows of gravitational wave counterparts and gamma-ray bursts (GRBs), and infer properties of the explosion and its environment. Such analysis can also aid in the rapid identification of kilonovae superimposed on the standard afterglow signal \citep[e.g.][]{Troja2023}. To achieve these aims, GWAPA provides a data repository for GW and GRB afterglows and a web-based interactive access to the \texttt{afterglowpy} \texttt{Python} package \citep{Ryan20} and incorporate five jet structures and a spherical outflow\footnote{Referred to as a cocoon in earlier releases of \texttt{afterglowpy}.} model into this release of GWAPA.  GWAPA includes an intuitive interface powered by the \texttt{Bokeh} \texttt{Python} package \citep{bokeh} allowing interactive plotting and comparison of models and observed data.

\section{Webtool design} \label{sec:design}

\subsection{Backend}

GWAPA is built around the \texttt{afterglowpy} \texttt{Python} package \citep[\texttt{v0.7.3,}][]{Ryan20}. \texttt{Afterglowpy} uses the single shell approximation of \citet{vanEerten10} and \citet{vanEerten18}, which approximates the physics of the ejecta and forward shock as a single radially uniform fluid element. Jet spreading is also approximated and a trans-relativistic equation of state interpolates between the ultra-relativistic and non-relativistic regimes \citep{vanEerten13,Nava13}. \texttt{Afterglowpy} has been calibrated using the \texttt{BoxFit} code \citep{vanEerten12}, which derives afterglows from the results of high resolution 2D hydrodynamics simulations.

Traditionally, GRB jets are assumed to be `top hat' jets, with roughly constant energy across the jet and a precipitous decline at its edge. However, it is likely that jets have significantly more structure and \texttt{afterglowpy} includes several different possibilities. 
For instance, both Gaussian and power law structures are modelled. The energies of these jets varies with angle $\theta$ as 
\begin{equation}
    E(\theta) = E_0 \exp \left( - \frac{\theta^2}{2\theta_c^2} \right)
    \label{eq:gaussian}
\end{equation}
and
\begin{equation}
    E(\theta) = E_0 \left( 1 + \frac{\theta^2}{b\theta_c^2} \right)^{-b/2}
    \label{eq:plaw}
\end{equation}
respectively, where $\theta_c$ is the width and $b$ is the power law index. To model a structured jet, \texttt{afterglowpy} breaks it down into many top hat components. The properties of the blast wave for each of these components, primarily the radial position of the shock, dimensionless four-velocity and time-dependent jet opening angle, are described by a set of ordinary differential equations. These are numerically integrated and solved and the top hat components are summed over to derive the resultant afterglow of the GRB. 

We used \texttt{afterglowpy} to generate a grid of predefined models for five jet structures (top hat, Gaussian, Gaussian with core, smooth power law, power law with core) and a spherical outflow model. For each structure, we generated the model data at each point in a parameter grid covering the diverse population of GRBs, for example, $E_{\rm \gamma, iso}$ from $10^{49}$ to $10^{55}$ erg and observer angles from 0\degr~to $\sim$45\degr. For the spherical outflow, a different parameter grid was used due to the nature of the physics involved. We note that these parameter grids are somewhat coarse, again to limit computational load, however the grid is still fine enough to allow inference of GRB properties to a reasonable degree of precision and identify significant outliers to typical afterglow behaviour. Interactive tools, described in Section \ref{sec:ui}, are used to select a model to be plotted and compared with observed data. This allows the user to infer properties of a GRB, again while minimising computational load.

\subsection{User interface}
\label{sec:ui}

    \begin{figure}
\plotone{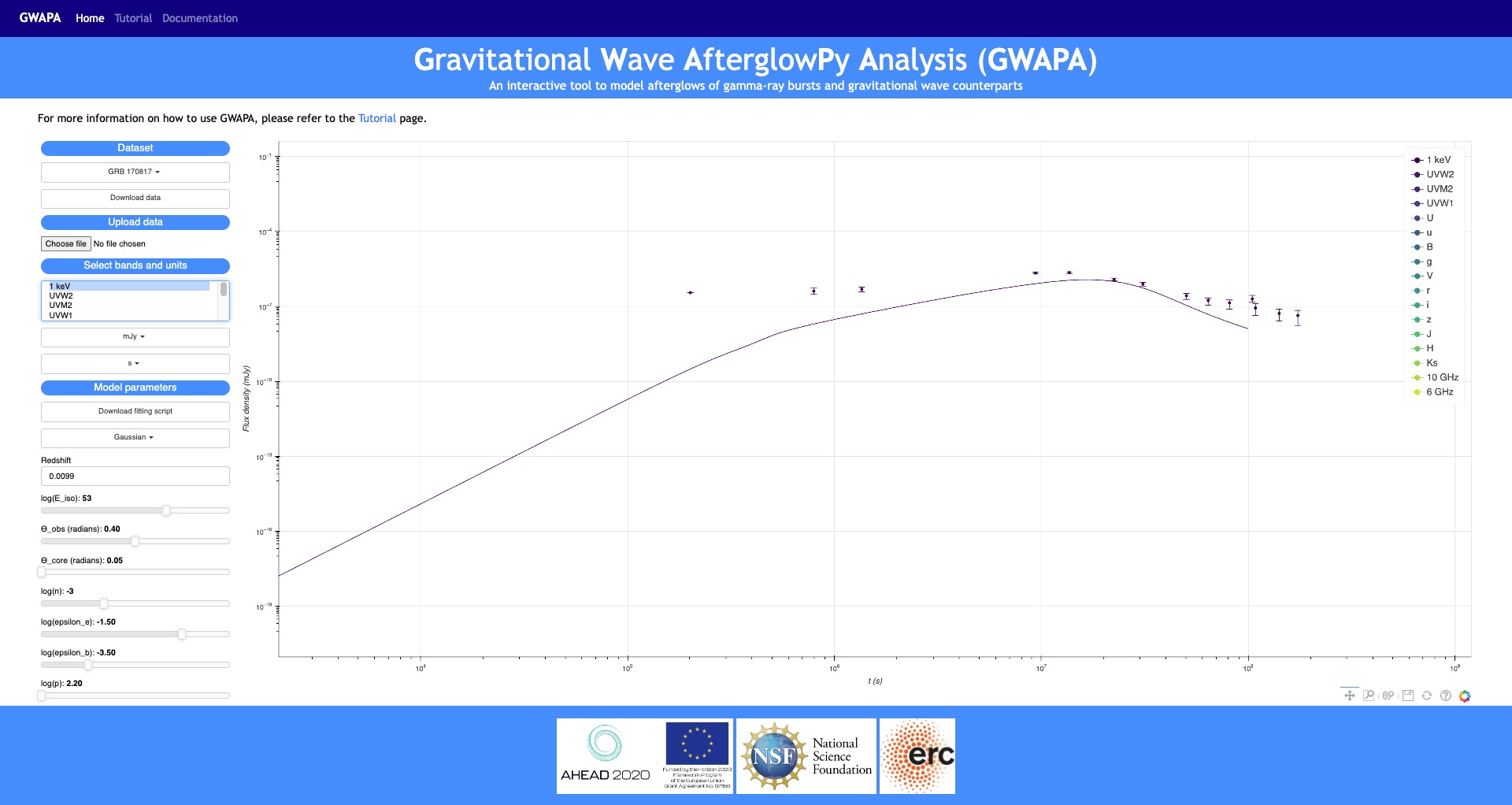}
\caption{The user interface for GWAPA, built using the \texttt{Bokeh} package. 
\label{fig:ui}}
\end{figure}

The user facing portion of the GWAPA webtool was developed using \texttt{Bokeh v2.4.3}, a \texttt{Python} package designed to provide exactly the type of interactive tool we desired \citep{bokeh}. \texttt{Bokeh} provides both a dynamic plotting package and a selection of `widgets', tools to control and interact with plots, which we have used to build a comprehensive user interface for GWAPA.

We show the user interface for GWAPA in Figure \ref{fig:ui}. It allows users to select a GRB from a small sample of examples to be plotted and downloaded, meaning GWAPA can also act as a data repository. While there are currently not plans to allow public uploads to this repository, it is a possibility that could be explored in future.. Alternatively, users can upload their own data which are automatically parsed assuming a specific format, details of which are given on GWAPA's Tutorial page. 

The bands to be plotted can also be selected and GWAPA provides a wide selection of options, ranging from X-ray to radio. All models include data for all bands allowing predictions for unobserved bands to be made with uploaded data. Both the flux density (Jy, mJy, $\mu$Jy or erg s$^{-1}$ cm$^{-2}$ Hz$^{-1}$) and time units (seconds or days) can also be changed for ease of interpretation. 

The user can select a jet structure and vary the parameters of the GRB and afterglow using \texttt{Bokeh} widgets which dynamically change to appropriate options, for instance, when the spherical outflow model is selected. The widgets are fixed to the ranges in the parameter grids and the relevant model is loaded and plotted to be compared with the selected data. The redshift is free to vary to any value and GWAPA assumes a Planck cosmology when adjusting model fluxes \citep{Planck18}. The user can then vary parameters appropriately to approximate the behaviour seen in their data.

A dynamically generated \texttt{Python} fitting script can also be downloaded. This script uses the values inferred using GWAPA as an initial guess for a full Markov Chain Monte Carlo fitting procedure using the \texttt{emcee} module \citep{emcee} and allows more accurate inference of the final parameters. This script is designed only as a starting point, and further refinement is likely required to produce publishable results. Nevertheless, it provides easy access to afterglow fitting with \texttt{afterglowpy}.

Finally, \texttt{Bokeh} includes some additional tools to interact with the plot, for instance, to zoom or to download and save the plot.

\section{Conclusion}

We have presented and described the first release of the Gravitational Wave AfterglowPy Analysis (GWAPA) webtool, designed to provide the community with an easy to use tool for rapid initial analysis of GRB afterglows. This release of GWAPA is now live and available at \url{https://gwapa.web.roma2.infn.it/}.

\begin{acknowledgments}
We thank Geoff Ryan for invaluable insight and discussion. 
This work was funded by the European Union’s Horizon 2020 Programme under the AHEAD2020 project (grant agreement number 871158),  by the European Research Council through the Consolidator grant BHianca (grant agreement ID 101002761) and  by the National Science Foundation (under award number 2108950).
\end{acknowledgments}

%

\vspace{5mm}


\software{afterglowpy \citep{Ryan20},  
          Astropy \citep{astropy13,astropy18,astropy22}, Bokeh \citep{bokeh}, Django \citep{django}, emcee \citep{emcee}
          }


\bibliography{gwapa}{}
\bibliographystyle{aasjournal}



\end{document}